%% file: aaskaii_template.tex
\documentclass[a4paper,11pt]{article}
\usepackage{aaskaiid}
\usepackage{orcidlink}
\newcommand\fd{\hbox{$.\!\!^{\reset@font\romn d}$}}
\newcommand\fh{\hbox{$.\!\!^{\reset@font\romn h}$}}
\newcommand\fm{\hbox{$.\!\!^{\reset@font\romn m}$}}
\newcommand\fs{\hbox{$.\!\!^{\reset@font\romn s}$}}

\newcommand\fp{\hbox{$.\!\!^{\reset@font\reset@font\scriptscriptstyle\romn p}$}}

\newcommand\arcsec{\hbox{$^{\prime\prime}$}}
\definecolor{mygray}{gray}{0.6}

\usepackage{booktabs,siunitx}

\title{Star Formation and Accretion in Nearby Galaxies}
\ShortTitle{SF \& Accretion in Nearby Galaxies}
\author[1]{J. Moldon\orcidlink{0000-0002-8079-7608}}
\ShortName{Moldon et al.} 
\emailAdd{jmoldon@iaa.es}

\author[1]{A. Alberdi\orcidlink{0000-0002-9371-1033}}
\author[1]{M. Perez-Torres\orcidlink{0000-0001-5654-0266}}
\author[1,2]{G. Lucatelli\orcidlink{0000-0002-2410-1776}}
\author[2]{R. Beswick\orcidlink{0000-0002-5544-2354}}
\author[3]{M. Sargent\orcidlink{0000-0003-1033-9684}}
\author[4]{E. Brinks\orcidlink{0000-0002-7758-9699}}
\author[5]{R. D. Baldi\orcidlink{0000-0002-1824-0411}}
\author[6]{S. Dey\orcidlink{0000-0002-4679-0525}}
\author[7]{F. S. Tabatabaei\orcidlink{0000-0002-0377-0970}}
\author[8]{K. Rubinur\orcidlink{0000-0001-5574-5104}}
\author[9]{N. Seymour\orcidlink{0000-0003-3506-5536}}
\author[10]{M. Pandey-Pommier\orcidlink{0000-0001-5829-1099}}
\author[11]{C. Bot\orcidlink{0000-0001-6118-2985}}

\affiliation[1]{Instituto de Astrof\'isica de Andaluc\'ia (IAA-CSIC), Glorieta de la Astronom\'ia s/n, 18008 Granada, Spain}
\affiliation[2]{Jodrell Bank Centre for Astrophysics, Department of Physics and Astronomy, The University of Manchester, Alan Turing Building, Oxford Road, Manchester M13 9PL, UK}
\affiliation[3]{\'Ecole Polytechnique F\'ed\'erale de Lausanne (EPFL), Station 3, CH-1015 Lausanne, Switzerland}
\affiliation[4]{Centre for Astrophysics Research, University of Hertfordshire, College Lane, Hatfield AL10 9AB, UK}
\affiliation[5]{INAF -- Instituto di Radioastronomia, via Gobetti 101, 40129 Bologna, Italy}
\affiliation[6]{National Centre for Nuclear Research (NCBJ), ul. Pasteura 7, 02-093 Warsaw, Poland}
\affiliation[7]{School of Astronomy, Institute for Research in Fundamental Sciences (IPM), PO Box 19395-5531, Tehran, Iran}
\affiliation[8]{Institute of Theoretical Astrophysics, University of Oslo, P.O. Box 1029 Blindern, N-0315 Oslo, Norway}
\affiliation[9]{ICRAR--Curtin University, Brodie Hall Building, 1 Turner Avenue, Technology Park, Bentley, WA 6102, Australia}
\affiliation[10]{Universit\'e Catholique de Lyon (UCLy), Campus Saint-Paul, 10 Place des Archives, 69288 Lyon Cedex 02, France}
\affiliation[11]{Observatoire astronomique de Strasbourg, Universit\'e de Strasbourg, 11 rue de l’Universit\'e, 67000 Strasbourg, France}

\abstract{The appearance of galaxies is strongly dominated by two physical phenomena: star formation and accretion of material onto compact objects, primarily supermassive black holes. Nearby galaxies offer a unique window to study these processes in detail and with high spatial resolution. The $\mu$Jy sensitivity, sub-arcsecond angular resolution and broadband coverage provided by SKA AA4 configuration will enable us to separate morphologically and spectrally the contribution from SMBH accretion and star formation processes, from the most compact nuclear region out to the most diffuse, galaxy-wide components. Disentangling thermal and non-thermal emission using multi-scale spectral index maps (traced by SKA-Mid) and absorption processes (traced mainly by SKA-Low) will allow us to account for the present and past star-formation rate, the AGN activity and the ISM properties. SKA observations of a wide range of galaxy types in the nearby Universe, including quiescent, AGN-dominated and starburst galaxies, will provide a broad view of how star formation and accretion regulate galaxy evolution, while time-domain and spectral-variability information further isolate compact accretion and transient phenomena.}
\begin{document}
\maketitle

\section{Introduction \& Scientific Motivation}
Star formation drives the buildup of stellar mass in galaxies, while accretion onto black holes powers some of the most energetic phenomena in the Universe. The interaction between these processes is central to explaining the observed galaxy luminosity function and how it is shaped by feedback processes. Radio observations provide a unique view, unaffected by dust, of both star formation and accretion. The Square Kilometre Array (SKA) in stage delivery AA4 will transform this field through its unprecedented combination of $\mu$Jy sensitivity, sub-arcsecond resolution, and continuous frequency coverage from 50~MHz to 15~GHz. These capabilities will enable detailed studies of nearby galaxies, from individual giant molecular clouds ($\sim10$~pc) in the nearest systems to complete samples within 200~Mpc \citep{beswick2015}.

The key motivation is to determine how star formation and accretion jointly regulate galaxy growth across environments and to link radio continuum diagnostics to the physical state of the interstellar medium, establishing the empirical framework for interpreting distant systems. Nearby galaxies allow us to map the interplay between cosmic rays, magnetic fields, and feedback processes that control radio emission efficiency.

Radio continuum emission provides simultaneous access to multiple physical components within a single observational framework \citep{condon1992,murphy2011}. Thermal free-free emission from H~II regions traces current ($\lesssim10$~Myr) massive star formation with minimal dependence on dust or metallicity \citep{murphy2011,kennicutt2012}. Non-thermal synchrotron emission probes star formation over longer timescales ($\sim30-100$~Myr) while constraining cosmic-ray transport, magnetic fields, and feedback coupling, making it indispensable for extending radio-SFR relations to dust-obscured galaxies where only the non-thermal component remains detectable \citep{heesen2019,tabatabaei2017sed}. Compact, typically flat-spectrum sources trace accreting black holes and provide diagnostics of jet activity across the full mass range from stellar remnants to supermassive black holes \citep{merloni2003}.

The SKA's wide frequency coverage enables diagnostics not accessible to current arrays. Low frequencies (SKA-Low: 50--350~MHz) probe synchrotron emission from long-lived cosmic-ray electrons, mapping extended halos \citep{heesen2009}, and can reveal low-frequency turnovers from free-free absorption in compact or embedded star-forming regions \citep{clemens2010,varenius2016}. High frequencies (SKA-Mid Band~5a-b: 4.6--15.4~GHz) increase the thermal free-free fraction and, with spectral decomposition, provide dust-insensitive SFR estimates while achieving the resolution required to separate nuclear AGN from surrounding star formation \citep{murphy2012,linden2020,herrero-illana2017}. High angular resolution imaging of $\sim0.5\arcsec$ (corresponding to physical scales of $\approx 50-500$\,pc at distances of 20--200\,Mpc) will resolve galaxies into compact radio populations (AGN, H~II regions, super star clusters, X-ray binaries, and supernova remnants) providing an unobscured census of massive star formation and accretion \citep{perez-torres2021,muxlow2020}. At a few~$\mu$Jy sensitivity, SKA surveys should detect long-lived radio supernova remnants within several tens of Mpc, directly tracing high-mass ($M>8\,M_\odot$) star formation and constraining the high-mass end of the IMF \citep{beswick2015}. On larger scales, SKA imaging of diffuse emission will separate synchrotron and thermal components through broadband spectral decomposition \citep{galvin2018,tabatabaei2022,dey2022}, Lucatelli et al. (in preparation). This multi-scale approach, from compact sources to galaxy-wide halos, connects local star-forming regions with global radio properties, enabling robust, environment-dependent calibrations anchored in nearby galaxies. The thermal emission provides a measure of massive star formation, whereas the non-thermal, synchrotron, component traces cosmic-ray and magnetic-field processes; together they establish a local reference for interpreting unresolved high-redshift galaxies. Applications to high-redshift radio-continuum measurements and the cosmic star-formation history are discussed in this book by \citet{Algera01.2026.SKA,FangxiaAn01.2026.SKA}.

This chapter outlines the physical framework, methodologies, and observational strategies that exploit the SKA's capabilities to disentangle star formation and accretion and to address fundamental questions in galaxy evolution.


\section{Physical Processes \& Observational Framework}

\subsection{Spectral energy distribution components and diagnostics}
\label{sec:sed_components}

\begin{figure}[!tb]
\center
\includegraphics[width=0.9\textwidth]{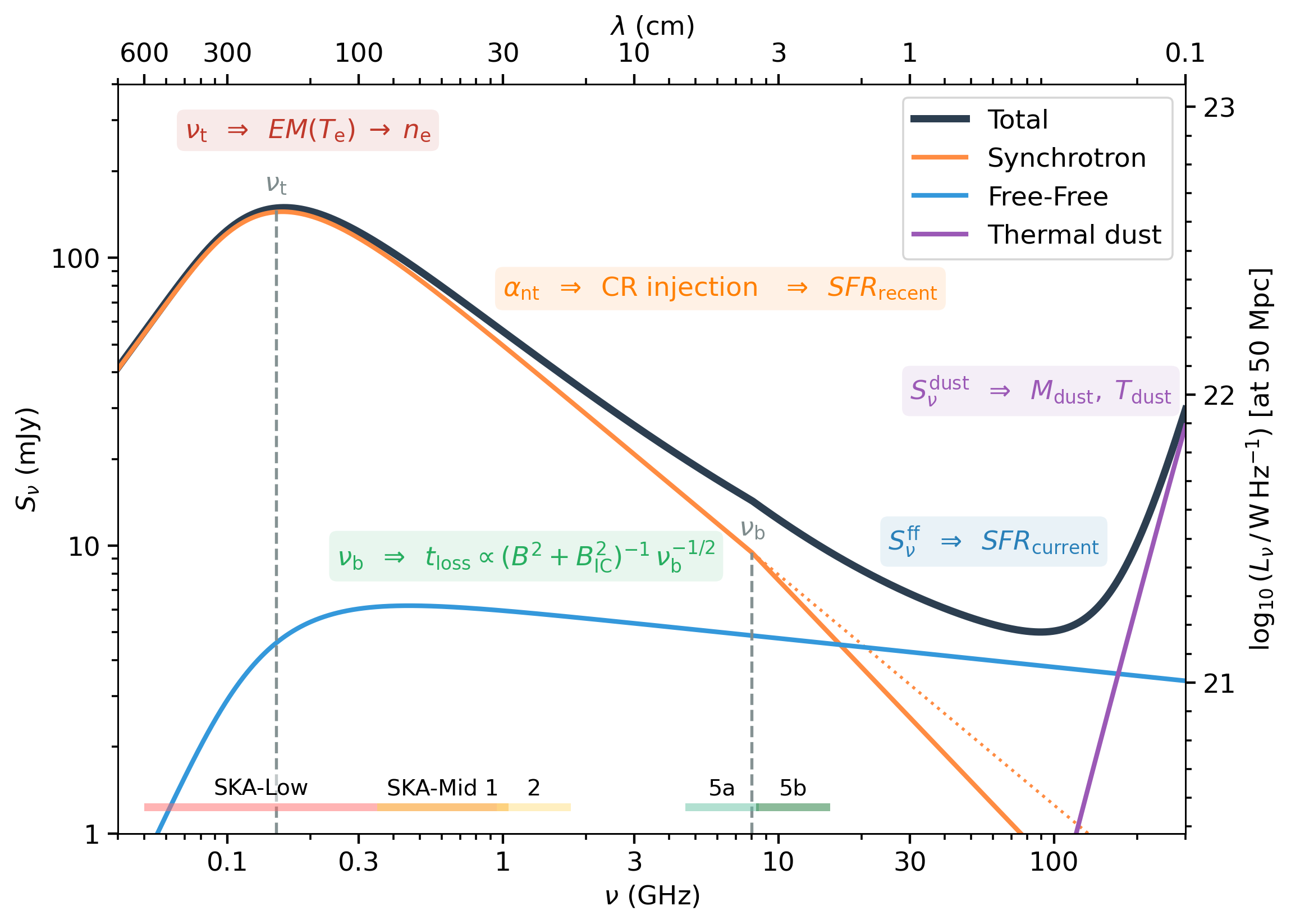}
\caption{\label{Fig:sed} Broadband radio-to-mm spectral energy distribution of a nearby star-forming galaxy, showing synchrotron (orange), free-free (blue), and thermal dust (violet) components, with total emission in black. Vertical dashed lines mark the free-free absorption turnover ($\nu_{\mathrm{t}}$) and synchrotron cooling break ($\nu_{\mathrm{b}}$); shaded bands indicate the SKA-AA4 frequency coverage. Key diagnostics include electron density, cosmic-ray injection, cooling timescales, and star-formation rate. The $\sim$1\,GHz range constrains the synchrotron slope, whereas 5--15\,GHz traces the transition to thermal emission. The model shown excludes any AGN or jet-related contribution, such as a flat variable core or a compact jet with a steeper spectrum.}
\end{figure}

The radio spectral energy distribution (SED) of galaxies encodes the balance between thermal and non-thermal emission mechanisms. In the SKA frequency range, its shape provides a direct means to quantify the relative contributions of free-free and synchrotron processes, and to trace the transport and energy losses of cosmic rays. Characterizing these components forms the basis for separating star formation and accretion-related emission and for deriving physical parameters such as magnetic-field strength, ionized-gas content, and recent star-formation history. The broader role of thermal and non-thermal processes in the ISM and IGM is discussed by \citet{Tabatabaei01.2026.SKA}. In Fig.~\ref{Fig:sed} we show a sketch of the main components of emission associated with star formation.

\textbf{Thermal free-free emission} originates from H~II regions surrounding young, massive stars and is directly proportional to the photoionization rate \citep{rubin1968,osterbrock2006}. This component provides a tracer of recent star formation activity with a nearly flat spectrum, where $\alpha_{\rm th} \approx -0.1$ for $S_\nu \propto \nu^\alpha$ \citep{niklas1997,murphy2011,murphy2012}. In AGN, thermal free-free emission may also arise from photoionized gas in the inner nucleus, including disk winds and AGN-driven outflows, and from compact coronal plasma in radio-quiet systems \citep[see][]{panessa2019}.

\textbf{Non-thermal synchrotron emission} arises from relativistic electrons accelerated in supernova remnants and propagating through galactic magnetic fields \citep{condon1992}. Its spectral slope ($\alpha \approx -0.8$ in normal star-forming galaxies) reflects the energy distribution of cosmic rays shaped by diffuse shock acceleration and subsequent radiative and escape losses \citep{longair2011}. In AGN, synchrotron radiation originates in jets and compact cores, where optical-depth effects can flatten the spectrum ($\alpha \gtrsim -0.3$) or, when unresolved jet components dominate, yield steep integrated spectra, $\alpha \approx -0.7$ \citep{baldi2021}.

SKA-AA4's frequency coverage enables spectral diagnostics across three key regimes. \textbf{Low frequencies (SKA-Low + Band 1, 50--1050 MHz)} are in general dominated by synchrotron emission with negligible thermal contamination, providing clean studies of cosmic-ray populations and magnetic fields \citep{dey2024,fangxia2024}. Extended synchrotron halos trace cosmic-ray transport on kiloparsec scales, while spectral-index maps separate steep star-forming emission from flatter compact nuclear sources and possibly unresolved AGN jets (Lucatelli et al., in preparation). Free-free absorption at these frequencies reveals deeply embedded star formation and constrains the structure of the ionized medium \citep{clemens2010,varenius2016,gajovic2025}, while synchrotron self-absorption can shape the low-frequency turnover in compact AGN cores. In addition, compact symmetric objects and GHz-peaked spectrum sources often show strong free-free absorption signatures, further modifying their low-frequency spectra.

\textbf{Mid frequencies (Bands 2+5a, 950--8500 MHz)} exhibit complex interplay between thermal and non-thermal components, enabling separation through spectral fitting. The transition from synchrotron-dominated to thermal-dominated emission varies with star formation surface density rate. Spectral curvature measurements across these bands break degeneracies, i.e. they distinguish between different combinations of cosmic-ray transport and energy-loss processes that would otherwise produce similar spectral slopes \citep{tabatabaei2017sed}.

\textbf{High frequencies (Band~5b, 8.3--15.4~GHz)} generally have enhanced thermal free-free contributions, with thermal fractions, $f_{\rm th}$, often exceeding 50\% in star-forming regions and reaching high values in compact H~II regions \citep{tabatabaei2007,murphy2012,linden2020}. The exact fraction remains environment-dependent: nearby non-AGN star-forming complexes can show median 33~GHz thermal fractions of order $f_{\rm th}\simeq0.7$--0.8, whereas LIRG nuclear-ring regions on $\sim100$--300~pc scales can remain below 50\% \citep{murphy2012,song2021}. Thermal-fraction mapping isolates free-free emission to obtain dust-independent SFR surface densities and, combined with non-thermal maps, constrains the coupling between magnetic fields and star formation. Sub-arcsecond imaging at these frequencies further separates compact flat-spectrum nuclear sources from extended star-forming complexes \citep{muxlow2020}.

\subsection{Cosmic ray transport and magnetic field physics}
\label{sec:cr_transport}

The previous section described the spectral signatures of emission components, understanding their spatial distribution requires accounting for cosmic-ray transport and magnetic-field physics.

The transport of cosmic ray (CR) electrons governs how non-thermal radio emission relates to recent star formation on resolved (sub-kpc) scales. In galaxy disks, CRs propagate away from their supernova acceleration sites by a combination of diffusion in turbulent magnetic fields and large-scale advection in galactic winds. Effective propagation lengths of order $\sim$kpc at GHz frequencies smooth the radio maps relative to instantaneous SFR tracers \citep{braun2007,murphy2008,heesen2014}.

Whether galaxies behave as electron calorimeters (most CR-electron energy radiated before escape) or allow significant CR leakage depends on magnetic fields, radiation fields, gas density, and wind speeds. Low–surface-density disks tend toward escape-dominated regimes, while dense starbursts approach calorimetry \citep{lacki2010,heesen2019}. This is set by the balance between the radiative loss timescale $t_{\rm loss}=E/|\mathrm{d}E/\mathrm{d}t|_{\rm loss}$ (synchrotron, inverse-Compton, bremsstrahlung, ionization) and the escape timescale $t_{\rm esc}$, given by diffusion ($t_{\rm esc}\simeq h^2/2D$, for scale height $h$ and diffusion coefficient $D$) or advection ($t_{\rm esc}\simeq h/v_{\rm adv}$  for wind speed $v_{\rm adv}$). Calorimetry occurs when $t_{\rm loss}\lesssim t_{\rm esc}$; efficient CR escape when $t_{\rm loss}\gg t_{\rm esc}$.

Broadband radio spectra provide diagnostics of cosmic-ray aging and transport. Curvature between low (SKA-Low + Band~1) and mid frequencies (Bands~2+5a) reflects the competition between synchrotron and inverse-Compton cooling versus ionization and bremsstrahlung losses, while spatial variations in spectral index across galaxy disks trace the propagation of CR electrons away from their star-forming acceleration sites \citep{tabatabaei2022}. Edge-on galaxies reveal the vertical dimension of CR transport through radio halos with scale heights of a few kiloparsecs. The frequency dependence of halo morphology and vertical spectral-index profiles distinguishes advection- from diffusion-dominated transport: large halos with gently steepening spectra indicate advection with wind speeds of order the escape velocity, whereas smaller halos with strong near-plane steepening favor diffusion \citep{heesen2016,schmidt2019}.

Magnetic fields fundamentally control the non-thermal radio emission through their regulation of cosmic ray transport, energy losses, and synchrotron radiation. Observations reveal a correlation between magnetic field strength and star formation surface density, $B \propto \Sigma_{\rm SFR}^{n}$ with $n \approx 0.3$--0.4 \citep{heesen2014,beck2015,tabatabaei2017sed}, consistent with turbulent dynamo amplification driven by supernova feedback. Typical equipartition field estimates range from a few to $\sim10~\mu$G in galaxy disks to $\sim50$--$100~\mu$G in central starburst regions \citep{beck2015}.

Synchrotron emission is intrinsically polarized, and because thermal free-free radiation is essentially unpolarized, SKA polarimetric imaging provides a strong constraint on the non-thermal component, although the separation is not fully model-independent because the observed polarization depends on magnetic-field ordering, Faraday effects, and viewing geometry \citep{beck2015}. SKA's broad frequency coverage will enable rotation measure synthesis to map three-dimensional magnetic field structure, where the rotation measure ${\rm RM} \propto \int n_e B_\parallel dl$ traces the line-of-sight magnetic field weighted by electron density \citep{brentjens2005,osullivan2012}. For nearby spiral star-forming galaxies at centimetre wavelengths, typical integrated polarization fractions are $p_{\rm obs}\sim 1$--$4\,\%$, with a range extending to $\sim 18\,\%$ at 4.8\,GHz \citep{stil2009}.


\section{Star Formation in Nearby Galaxies}

\subsection{Resolved SFR mapping}
\label{sec:sfr_mapping}

A central goal of SKA continuum studies is to obtain dust-unbiased maps of the current star-formation rate (SFR) from parsec to kiloparsec scales by isolating thermal free-free emission from the non-thermal synchrotron background. In the high-frequency SKA-Mid bands (5a/5b; 4.6--15.4~GHz), the thermal fraction rises rapidly, often exceeding 50\% in compact H\,\textsc{ii} regions, making free-free continuum a direct tracer of the ionizing photon rate from massive O stars \citep{murphy2011}.

The free-free luminosity $L_\nu^{\rm ff}$ scales with SFR as ${\rm SFR} \propto T_e^{0.45}\,\nu^{0.1}\,L_\nu^{\rm ff}$, yielding a calibration independent of dust and only dependent on the electron temperature $T_e$ \citep{murphy2012,kennicutt2012}. The SKA's broad bandwidth and high dynamic range enable per-pixel SED fitting that separates thermal and synchrotron components over matched beams, producing resolved maps of thermal fraction, spectral slope, and curvature. These maps identify the youngest H\,\textsc{ii} regions through elevated thermal fractions and reveal spatial variations in non-thermal emission linked to local star-formation intensity and cosmic-ray transport. Sub-arcsecond imaging with SKA-Mid recovers these quantities on cloud and association scales ($\sim$10--50~pc).

Cross-validation of thermal radio component against dust-corrected optical or UV tracers (e.g. FUV+IR or H$\alpha$+IR) provides an external calibration check and quantifies dust-related biases. On the other hand, resolved comparisons also reveal environment-dependent radio efficiency, i.e. the synchrotron luminosity per unit SFR. Low-metallicity dwarfs and outer disks fall below the canonical radio continuum (RC)--SFR calibration due to cosmic-ray escape and weaker magnetic fields \citep{schleicher2016}, whereas compact starbursts approach calorimetric conditions with enhanced non-thermal output \citep{lacki2010}.

\subsection{Compact populations for direct SFR determination}
\label{sec:stellar_populations}

Radio continuum and hydrogen radio recombination lines (RRLs) provide dust-insensitive constraints on the massive stellar content and its recent evolution, with recent work showing that high-frequency radio continuum and RRLs are among the most reliable SFR tracers \citep[e.g.][]{humire2025}. At SKA-Mid frequencies, especially Band~5b (8.3--15.4\,GHz), RRLs (e.g. H91$\alpha$, H92$\alpha$) fall within the bandpass while the continuum becomes increasingly thermal, enabling joint line+continuum diagnostics of $T_e$, $n_e$, and $Q({\rm H}^0)$ on sub-kpc scales (see Section~\ref{sec:sed_components}; \citealt{anantharamaiah1993,perez-torres2021}). Young massive clusters and super star clusters (SSCs) are laboratories for early feedback. Ultra-dense H\,II regions yield partially optically thick free-free spectra with rising or flat indices ($\alpha \gtrsim 0$) turning over to optically thin at higher frequencies. Extreme examples such as the NGC~5253 ``radio supernebula'' host $\lesssim$pc-scale high-brightness thermal sources requiring thousands of O\,stars \citep{turner2004}.

Core-collapse supernovae (CCSNe) and supernova remnants (SNRs) trace the formation and death of massive stars on 10--100\,Myr timescales and complement the instantaneous RRL/free-free view. CCSN radio emission arises from shock interaction with circumstellar material, with light curves and spectra encoding mass-loss rates and magnetic-field amplification; SNR populations trace the recent CCSN rate and correlate with host properties, constraining the high-mass end of recent star formation \citep{perez-torres2021}. In compact starbursts, SKA sensitivity and cadence will yield extinction-free CCSN rates for direct comparison with SFR estimates. Together, these diagnostics constrain the high-mass IMF: free-free/RRLs capture the ionizing photon output, while CCSN counts trace the deaths of stars with $M \gtrsim 8\,M_\odot$. Resolved measurements across diverse environments will test whether the IMF used in SFR calibrations is universal or varies with metallicity, density, or star-formation intensity. These independent constraints on massive star formation timescales inform the environment-dependent calibrations discussed in Section~\ref{sec:env_scaling}.
\subsection{Environment dependencies and radio-SFR calibration}
\label{sec:env_scaling}

The radio continuum--SFR relation varies with environment through cosmic-ray (CR) transport and magnetic-field coupling. In low surface-density systems (dwarfs, outer disks), weak fields and low gas densities promote rapid CR escape, reducing the non-thermal radio efficiency at fixed SFR and producing sublinear resolved RC--SFR slopes and radio deficits relative to dust- or H$\alpha$-based tracers \citep{heesen2014}. Low metallicity enhances these deficits by limiting dust cooling and magnetic-field amplification and by increasing ISM porosity, which shortens CR electron residence times \citep{schleicher2016}. At the opposite extreme, compact nuclear starbursts approach electron calorimetry: high gas and radiation energy densities and strong fields shorten synchrotron lifetimes, boosting radio output per unit SFR. This behavior corresponds to the regime where $t_{\rm loss} \lesssim t_{\rm esc}$ as discussed in Section~\ref{sec:cr_transport}.

A practical radio--SFR prescription must incorporate both components and their environmental modulation. We adopt a two-component description per resolution element and frequency $\nu$,

\begin{equation}
L_\nu = L^\mathrm{ff}_\nu (Q, T_\mathrm{e}) + L^\mathrm{nt}_\nu (\alpha_\mathrm{nt}, \eta_\mathrm{radio}(\nu | \Sigma_\mathrm{SFR}, Z, B, D, v_\mathrm{w})) \exp[-\tau_\mathrm{ff}(\nu)],
\end{equation}
where the exponential term accounts for free-free absorption affecting the background synchrotron emission,  $L_\nu^{\rm ff}$ traces the ionizing-photon production rate $Q$ with weak dependence on the electron temperature $T_{\rm e}$. The non-thermal term depends on the synchrotron slope $\alpha_{\rm nt}$ and an effective radio efficiency $\eta_{\rm radio}$ that varies with star-formation surface density $\Sigma_{\rm SFR}$, metallicity $Z$, magnetic-field strength $B$, gas density $D$, and wind speed $v_w$ through their influence on CR injection, cooling, and escape \citep{lacki2010}. The free-free absorption factor $\exp[-\tau_{\rm ff}(\nu)]$ becomes significant at low frequencies in deeply embedded regions (see Section~\ref{sec:cr_transport}).

\textbf{Frequency-dependent prescriptions.} As established in Section~\ref{sec:sed_components}, each SKA frequency regime provides distinct advantages for SFR measurement. At high frequencies, thermal fractions are largest and SFR maps follow from $L_\nu^{\rm ff}$. The commonly used calibration ${\rm SFR} \propto T_e^{0.45} \nu^{0.1} L_\nu^{\rm ff}$ applies to recent SF and is preferred in heterogeneous environments because it is insensitive to CR escape \citep{murphy2011,kennicutt2012}. At intermediate frequencies, two-component per-pixel fits recover thermal fraction $f_{\rm th}$, non-thermal index $\alpha_{\rm nt}$, and curvature, enabling SFRs from the thermal term and CR transport constraints through the inferred radio efficiency \citep{heesen2014,tabatabaei2017sed}. Low-frequency data require high-frequency anchoring and corrections for free-free absorption and curvature; they constrain long-lived CRs and the environment dependence of the radio efficiency when combined with higher-frequency observations.


\section{Accretion Across Mass Scales in Nearby Galaxies}

\subsection{SMBHs and low-luminosity AGN}
\label{sec:smbh_llagn}

A complete census of nuclear radio activity in nearby galaxies requires separating compact, accretion-powered cores from star-formation emission using sub-arcsecond imaging and broad frequency leverage. Low-luminosity AGN (LLAGN) are identified through flat to near-flat spectra at GHz frequencies ($\alpha \gtrsim -0.3$), with inverted spectra corresponding to $\alpha > 0$, high brightness temperature, morphological compactness, and variability.

The brightness temperature criterion is particularly powerful:
\begin{equation}
T_b = \frac{c^2}{2k_B\nu^2} \frac{S_\nu}{\Omega} \simeq 1.22 \times 10^{12} \frac{S_\nu[{\rm Jy}] \lambda^2[{\rm cm}^2]}{\theta_{\rm maj}[{\rm mas}] \theta_{\rm min}[{\rm mas}]} \, {\rm K}
\end{equation}
where $\theta_{\rm maj}$ and $\theta_{\rm min}$ are the deconvolved major and minor axes of the emitting region, and $T_b \gtrsim 10^5$--$10^6$\,K excludes thermal H\,II emission and supports a non-thermal AGN core \citep{morabito2022}. Combined with compact morphology and variability at the 10--50\% level on timescales of months to years, this yields robust, model-independent LLAGN selection.

In crowded, dusty nuclei (e.g. LIRGs/ULIRGs), VLBI-scale detections with $T_b \gg 10^6$ K cleanly isolate accretion-powered cores from clusters of young SNe/SNRs \citep{romero-canizares2012,perez-torres2021}. SKA-Mid sensitivity and resolution will extend such core detections to lower luminosities and larger distances than current surveys, including weak or radio-quiet AGN. The combination of arcsecond-scale morphology from SKA-Mid and milliarcsecond-scale structure from SKA-VLBI will provide unprecedented leverage for AGN identification. The broader connection between magnetized accretion disk--jet systems from stellar-mass to supermassive black holes is discussed in this book by \citet{Pathak01.2026.SKA}.

Integrated radio luminosities, preferably including resolved jet/lobe emission rather than unresolved cores alone, can be used as order-of-magnitude proxies for kinetic power through empirical radio-cavity/jet scalings, with large intrinsic scatter \citep{cavagnolo2010}. Host coupling can be evaluated by comparing core/jet maps with molecular gas tracers, where cavities or limb-brightened edges flag jet-ISM interaction on 10--1000\,pc scales \citep[e.g.][]{morganti2013}. When jets dominate over faint cores or remain unresolved on sub-parsec scales, the integrated spectrum can appear steep, making high-resolution imaging essential to separate nuclear emission from extended jet structures and to avoid misclassifying radio-quiet or low-luminosity AGN as star-forming sources.

\subsection{Sub-galactic accretion: stellar remnants and intermediate-mass objects}
\label{sec:subgalactic_accretion}

Compact accretors such as stellar-mass black holes and neutron stars in X-ray binaries (XRBs), ultra-luminous X-ray sources (ULXs), putative intermediate-mass black holes (IMBHs), and tidal disruption events (TDEs) can contaminate radio-based SF measurements if unrecognized. Their emission is compact, often flat, and in some cases inverted, at a few GHz when partially self-absorbed jets are present, and variable on timescales of days to years.

With SKA-Mid sensitivity and $\sim0.1$--$0.3''$ beams (50--300\,pc at 100--200\,Mpc), many such sources will be detected in high-SFR systems; screening relies on spectral slope, compactness, variability, and cross-matching to X-ray catalogs. In stellar-mass accretors, the low/hard state hosts steady, compact jets with flat or mildly inverted ($\alpha > 0$) spectra \citep{fender2001} and a tight radio--X-ray coupling \citep{gallo2003}. ULXs can further complicate interpretation, as they produce both compact radio cores and extended bubbles \citep[e.g.][]{kaaret2017,panurach2024}, and recent work shows that their radio emission can overlap the luminosity range of weak AGN.

TDEs in dusty merger systems supply time-domain nuclear radio emission unrelated to ongoing star formation, with mildly relativistic outflows that can persist for years \citep[e.g.][]{mattila2018,cendes2024}. Regular cadence imaging and spectral monitoring with SKA will be essential for distinguishing evolving TDE afterglows from steady LLAGN cores and thermal H\,II regions, with brightness temperature thresholds separating non-thermal jets and transients from H\,II regions. These observations will yield catalogs of sub-galactic accretors for demographic studies.


\section{Observational Methodology and Source Separation}

The observational program is designed to exploit the unique frequency, resolution, and sensitivity domain of the SKA. Current facilities such as JVLA, MeerKAT, e-MERLIN or LOFAR only partially sample this space. SKA-AA4 provides almost continuous coverage from 50 MHz to 15 GHz with high resolution imaging (up to tens of mas at the highest frequencies, plus VLBI capabilities) and $\mu$Jy sensitivity, enabling, in a single dataset, extinction-free SFR mapping from thermal emission and simultaneous constraints on non-thermal processes, absorption, and compact accretion signatures. This combination is essential to disentangle the physical drivers of the radio emission and to calibrate robust SFR and AGN diagnostics for galaxies in all environments.

\subsection{Multi-scale spectral decomposition framework}
\label{sec:spectral_decomposition}

Extracting robust physical diagnostics requires systematic separation of thermal and non-thermal components across spatial scales and frequency ranges. The framework employs coordinated broadband observations across SKA-Low and SKA-Mid with matched spatial resolution where feasible \citep[see][]{lucatelli2024}.

\textbf{Spectral fitting methodology.} Per-pixel spectral energy distribution fitting employs a two-component model with free-free and synchrotron emission, where Band 5b (8.3--15.4 GHz) anchors the thermal component and SKA-Low provides leverage on spectral curvature. The thermal component directly traces recent ($\lesssim 10$ Myr) massive star formation via ${\rm SFR} \propto T_e^{0.45} \nu^{0.1} L_\nu^{\rm th}$ \citep{murphy2011}, while the non-thermal component encodes cosmic ray transport and energy losses. Bayesian fitting propagates uncertainties through to derived quantities, enabling robust error estimates on thermal fraction, spectral index, and curvature parameters \cite{westcott2018}.

\textbf{Key observational diagnostics.} At low frequencies, free-free absorption with optical depth $\tau_{\rm ff} \propto \nu^{-2.1} T_{e}^{-1.35} {\rm EM}$ reveals deeply embedded star formation, with turnover frequencies constraining the emission measure and thus the density of the ionized gas \citep{clemens2010}. At frequencies below $\sim 200$--300\,MHz in dense star-forming environments, ionization losses and free-free absorption can produce spectral flattening or turnovers; the characteristic frequency depends on gas density, ionized-gas emission measure, magnetic and radiation fields, and source geometry \citep{lacki2010,grundy2025,gajovic2025}.

The synchrotron emissivity per unit volume is given by $j_\nu \propto N_0 B^{(\gamma+1)/2} \nu^{-(\gamma-1)/2}$, where $N_0$ is the cosmic ray electron density normalization, $B$ is the magnetic field strength, and $\gamma \approx 2.6$ is the electron energy spectral index \citep{longair2011}. For a typical electron spectrum, this yields $j_\nu \propto B^{1.8} \nu^{-0.8}$. Minimum energy arguments assuming equipartition between cosmic rays and magnetic fields yield equipartition field strength estimates $B_{\rm eq} \propto (L_{\rm syn,\nu}/V)^{2/7}$ where $L_{\rm syn,\nu}$ is the synchrotron luminosity at frequency $\nu$ (typically evaluated at 1.4\,GHz) and $V$ is the emitting volume \citep{beck2005}.

Polarimetric observations leverage the fact that synchrotron emission can reach $\simeq 70$--$75\,\%$ polarization in ordered fields while free-free emission is unpolarized \citep{beck2015}. In practice, the magnetic field in and around star-forming regions is highly turbulent, and beam depolarization strongly suppresses the polarized signal on these scales; the observed polarization therefore mainly traces the ordered large-scale field. The polarization fraction $p_{\rm obs} = p_0 \exp(-2\sigma_{\rm RM}^2 \lambda^4)$ decreases with wavelength due to Faraday dispersion, where $\sigma_{\rm RM}$ characterizes the rotation-measure distribution \citep{osullivan2012}. Multi-frequency polarimetric imaging enables rotation-measure synthesis to separate these effects and map the underlying magnetic-field structure \citep{brentjens2005}.

\subsection{AGN identification and separation}
\label{sec:agn_identification}

Building on the AGN diagnostics introduced in Section~\ref{sec:smbh_llagn}, we can establish a systematic identification framework. Robust AGN separation exploits the distinct properties of accretion-powered emission. At 50--200~Mpc, SKA-Mid achieves 25--100~pc resolution, sufficient to resolve star-forming complexes while AGN cores remain point-like. Using the brightness-temperature criterion from Equation~(2), cores with $T_b \gtrsim 10^6$ K exclude thermal processes. Adopting the convention $S_\nu \propto \nu^\alpha$, steep spectra have $\alpha < 0$, flat spectra have $\alpha \approx 0$, and inverted spectra have $\alpha > 0$. AGN cores exhibit flat or inverted spectra (typically $\alpha \gtrsim -0.3$, and inverted when $\alpha > 0$) due to synchrotron self-absorption, contrasting with steep star-formation spectra ($\alpha \approx -0.8$). Variability on month–year timescales with 10--50\% amplitudes provides additional confirmation.

Free–free absorption in compact ($<$10\,pc) nuclear starbursts is typically the main cause of flat or inverted spectra mimicking AGN cores, with synchrotron self-absorption contributing only in the most compact, high-brightness regions. Distinction requires resolved imaging: free–free–absorbed starburst emission is spatially extended on tens-to-hundreds of parsec scales, whereas genuine AGN cores remain compact and can reach high brightness temperatures \citep{barcos-munoz2015,romero-canizares2012,varenius2016}. Here, ``inverted'' denotes $\alpha > 0$ under the convention above. Probabilistic classification should combine spectral slope, compactness, brightness temperature, and variability.

\subsection{Direct supernova counting}
\label{sec:sn_methodology}

Direct detection and counting of core-collapse supernovae (CCSNe) provides an independent probe of recent massive star formation ($M \gtrsim 8\,M_\odot$). The CCSN rate links to SFR through $\Re = k_{\rm IMF}\,\mathrm{SFR}$, where $k_{\rm IMF}$ depends on the assumed IMF and is typically $(5$--$10)\times10^{-3}\,\mathrm{yr}^{-1}/(M_\odot\,\mathrm{yr}^{-1})$ for standard choices \citep{perez-torres2021}. High-cadence monitoring discovers new CCSNe and tracks their spectral evolution, yielding robust rate estimates even in optically opaque nuclei. VLBI resolves compact explosions from SNR complexes and AGN cores via brightness temperature and expansion measurements. Methodology and technical requirements are detailed in \citet{beswick2015}. Although IMF-sensitive in principle, current CCSN statistics in nearby starbursts (e.g. M82, with a measured rate of $\sim 0.1,\mathrm{yr^{-1}}$) remain too limited to distinguish between Salpeter-, Kroupa-, or mildly top-heavy IMFs, owing to small-number uncertainties and degeneracies with starburst age, extinction, and metallicity. The SKA will expand CCSN measurements to large, homogeneous samples, improving demographics and enabling tests for strong IMF deviations in extreme environments.



\section{Observational Program}

The three-tier SKA continuum program proposed here links resolved physics with population-level scaling relations. Deep, spectrally and morphologically resolved observations of nearby galaxies probe the low end of the stellar-mass and SFR distributions and resolve small-scale structures. Intermediate-depth mapping extends this analysis across diverse environments, testing the robustness of local calibrations, whereas wide-area surveys provide the statistical foundation to generalize these relations across the galaxy population. See Table~\ref{tab:obs_program} for some reference numbers.

\begin{table*}[t]
\centering
\caption{Three-tier continuum observing strategy for SKA-AA4. For each frequency band, the table lists the highest angular resolution ($\theta_{\max}$ with Briggs weighting with robust -2), the corresponding linear scale at each representative distance, and the 3$\sigma$ 1\,h flux-density sensitivity ($S_{\nu, 3\sigma}^{\rm 1h}$ assuming a tradeoff robust parameter of 0). The resulting $\log L_{\nu}$ values represent the corresponding 3$\sigma$ luminosity thresholds at 20, 50, and 200\,Mpc.}
\label{tab:obs_program}
\begin{tabular}{@{}lccccccccc@{}}
\toprule
 & & & &
\multicolumn{2}{c}{\textbf{Tier 1}} &
\multicolumn{2}{c}{\textbf{Tier 2}} &
\multicolumn{2}{c}{\textbf{Tier 3}} \\
 & & & &
\multicolumn{2}{c}{\footnotesize ($D<20$\,Mpc)} &
\multicolumn{2}{c}{\footnotesize ($D<50$\,Mpc)} &
\multicolumn{2}{c}{\footnotesize ($D<200$\,Mpc)} \\
\cmidrule(lr){5-6} \cmidrule(lr){7-8} \cmidrule(lr){9-10}

Band & $\nu_{\rm c}$ & $\theta_{\rm high}$ & $S_{\nu, 3\sigma}^{\rm 1h}$ & Scale & $\log L_{\nu}$ &  Scale & $\log L_{\nu}$ & Scale & $\log L_{\nu}$ \\
     & (GHz) & (") & ($\mu$Jy) & (pc) &  & (pc) &  & (pc)  \\
\midrule
Low  & 0.2  & 3.03 & 34.98 & 294  & 18.03 & 735  & 18.83 & 2938 & 20.03 \\
1    & 0.8  & 0.34 & 7.02  & 33   & 17.34 & 83   & 18.14 & 330  & 19.34 \\
2    & 1.3  & 0.21 & 3.42  & 20   & 17.02 & 51   & 17.82 & 200  & 19.02 \\
5a   & 6.6  & 0.04 & 2.11  & 3.9  & 16.84 & 9.7  & 17.64 & 39   & 18.84 \\
5b   & 11.9 & 0.02 & 2.53  & 1.9  & 16.93 & 4.9  & 17.73 & 20   & 18.93 \\
\bottomrule
\end{tabular}
\end{table*}

\textbf{Tier~1 (Resolved benchmarks).} 
A focused sample of nearby galaxies ($D \lesssim 20$\,Mpc; $\sim$5--10 systems) observed across key SKA frequency bands enables full spectral decomposition at sub-100\,pc scales with SKA-Mid. Targets span face-on and edge-on systems to probe spiral structure and vertical CR transport, matching PHANGS/THINGS/KINGFISH distances for direct synergy with molecular, ionized, and dust-gas diagnostics. SKA-Mid bands 2 and 5 (0.95--15.4\,GHz) provide the sensitivity and resolution for robust thermal/non-thermal separation. SKA-Low and Mid band~1 coverage can be drawn from survey products where they meet the required surface-brightness sensitivity and dedicated observations otherwise. A modest cadence of repeated observations supports variability studies of compact nuclear sources and enables detection and identification of new SNe and other transients.

\textbf{Tier~2 (Population diversity).}
A statistically meaningful sample to $D \lesssim 50$\,Mpc (several tens of galaxies) tests whether Tier~1 calibrations remain robust across stellar masses $>10^{7}\,M_\odot$, and SFRs $>0.01\,M_\odot\,{\rm yr}^{-1}$, extending down to dwarf irregular systems, and spanning morphologies from quiescent early-types through starbursts. SKA-Mid band~2 ($\sim$1.4\,GHz) delivers uniform sub-kpc resolution, enabling morphological separation of diffuse and compact emission, complemented by SKA-Low products and targeted band~5 follow-up. Targets are drawn from volume-limited samples with archival multi-wavelength coverage (optical IFU, FIR, HI) for pixel-by-pixel comparison. This tier quantifies dependencies on environment, metallicity, and gas conditions, and identifies outliers for deeper follow-up. Multi-epoch observations over months to years further allow demographic studies of variable and transient compact sources across diverse environments.

\begin{figure}[t]
\centering
\includegraphics[width=0.8\linewidth]{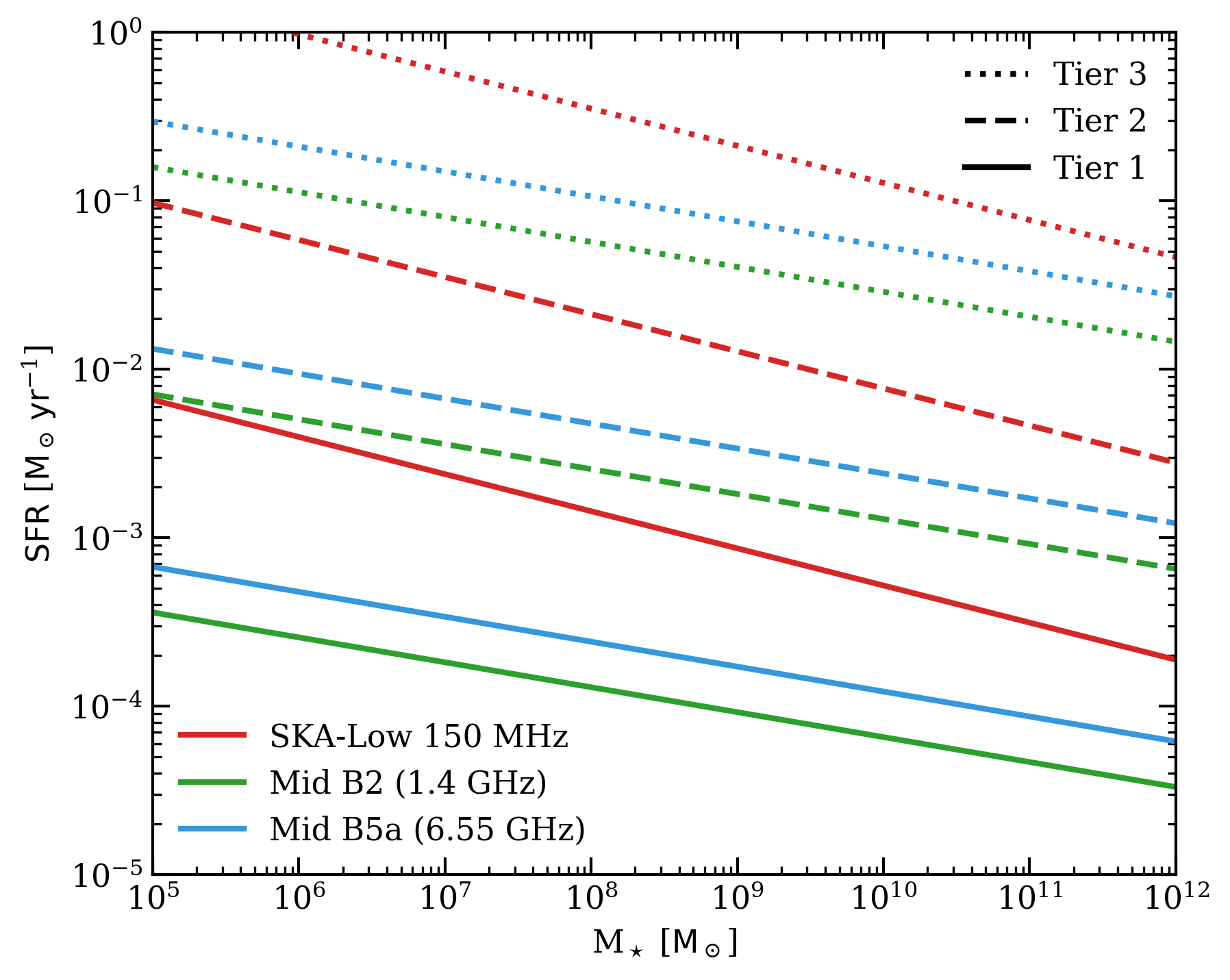}
\caption{SFR--$M_\star$ 3$\sigma$ detection threshold for the proposed observations, assuming 10, 5 and 0.5 hours for Tier 1, 2 and 3, respectively. For each band, curves (Tier 1/2/3: solid/dashed/dotted) show the SFR required at a given $M_\star$ for the expected total radio luminosity $L_\nu(\mathrm{SFR},M_\star)$ to equal the tier sensitivity at 20/50/200 Mpc. $L_\nu(\mathrm{SFR},M_\star)$ follows the 150 MHz calibration of Shenoy et al. (in preparation) for SKA-Low and the calibration of \citet{delvecchio2021} for SKA-Mid; the latter is scaled to 1.35 and 6.55 GHz using a fixed spectral index $\alpha=-0.7$ (the spectral index is not applied to Shenoy at 150 MHz).}
\label{Fig:sfr_mass_limits}
\end{figure}

\textbf{Tier~3 (Statistical anchor).}
A volume-limited sample to $D \lesssim 200$\,Mpc (hundreds to low thousands of galaxies) provides the statistical foundation for radio luminosity functions, the FIR--radio correlation, and calibration dependencies on stellar mass and sSFR. Tier~3 combines targeted SKA-Mid band~2 observations of benchmark galaxies ($\sim$100--200 objects filling parameter-space gaps) with commensal detections from wide-area SKA surveys. At these distances, band~2 achieves $\sim$200\,pc resolution, sufficient to distinguish nuclear from disk emission while remaining sensitive to SFR $\gtrsim 0.01-1\,M_\odot\,{\rm yr}^{-1}$ (Fig.~\ref{Fig:sfr_mass_limits}). This tier captures rare objects and enables cross-matching to deep multi-wavelength surveys (\emph{JWST}, \emph{Euclid}, LSST).

Figure~\ref{Fig:sfr_mass_limits} illustrates the nominal SFR--$M_\star$ detection thresholds for the three tiers. Standardized SKA pipelines will produce calibrated continuum and polarization maps, spectral-index and curvature maps, thermal-fraction maps from per-pixel SED decomposition (Section~\ref{sec:spectral_decomposition}), and compact-source catalogs with probabilistic AGN/star-formation classifications (Section~\ref{sec:agn_identification}). Tier~3 provides population-level quantities—radio luminosity functions, FIR--radio correlation parameters, and environment-dependent SFR calibrations (Section~\ref{sec:radio_scaling_populations}). For Tiers~1 and~2, repeated observations over months to years enable monitoring of compact variable sources (LLAGN, TDEs, SNe) and, with VLBI baselines, resolve proper motions on parsec scales.

\section{Key Scientific Outcomes and Legacy}

The SKA continuum program will deliver a comprehensive framework linking radio emission to star formation and accretion in nearby galaxies. Resolved observations down to $\sim$50\,pc will establish environment-dependent radio--SFR calibrations accounting for the transition from escape-dominated dwarfs to calorimetric starbursts (see Section~\ref{sec:cr_transport}), incorporating stellar-mass dependence, metallicity trends, and frequency-specific corrections. Broad-band spectra (150\,MHz--15\,GHz) will constrain dominant cosmic-ray loss mechanisms and determine calorimetric efficiency. Magnetic-field strengths from equipartition will test dynamo scaling relations and quantify coupling between magnetic energy and star formation.

Compact-source catalogs will provide extinction-free supernova and SNR rates, identify accretion-powered sources within star-forming nuclei, and track variability. Time-domain analysis will yield CCSN rates, transient phenomena, and LLAGN variability, enabling joint constraints on the high-mass IMF and accretion demographics.

\subsection{Radio luminosity scaling relations in galaxy populations}
\label{sec:radio_scaling_populations}

The FIRC links dust-reprocessed stellar light and non-thermal synchrotron emission over several orders of magnitude. Deep 150\,MHz LOFAR surveys find a super-linear relation with slope 1.37, median infrared-to-radio ratio $q_{\rm IR}$(150\,MHz) $\equiv \log_{10}(L_{\rm IR}/L_{150\,{\rm MHz}})$ = 2.14 $\pm$ 0.01, and scatter $\simeq$ 0.34 dex \citep{wang2019,smith2021}. Recent 150\,MHz studies reveal a secondary dependence on stellar mass: at fixed SFR, more massive galaxies are systematically more radio-luminous, following $L_{\rm radio}/{\rm SFR}\!\propto\!M_\star^{0.2-0.3}$ \citep{smith2021}, reflecting the transition from cosmic-ray escape in low-mass galaxies to calorimetry in massive systems.

Frequency-dependent effects are central to calibration. High frequencies (8--15\,GHz) trace thermal free-free emission insensitive to CR transport \citep{murphy2011,murphy2012}. Mid frequencies (1--8\,GHz) require two-component decomposition. Low frequencies (50--350\,MHz) are most affected by CR escape; at 150\,MHz, $L_{150~\rm MHz}=L_c\,\psi^{\beta}\!\left(M_\star/10^{10}\,M_\odot\right)^{\gamma}$ for SFR $\psi$ \citep{smith2021}. The 1.4\,GHz FIRC is mildly non-linear, with the infrared-to-radio ratio decreasing with radio luminosity \citep{molnar2021}. MIGHTEE shows reduced scatter when SFRs are averaged over 200--300\,Myr, consistent with CR electron lifetimes \citep{cook2024}.

The SKA's 50\,MHz--15\,GHz coverage, $\mu$Jy sensitivity, and sub-arcsecond resolution will enable full spectral decomposition and establish frequency- and mass-dependent calibrations. This provides the definitive local benchmark for high-redshift radio data and accurate SFR estimates from metal-poor dwarfs to ULIRGs. Population-level FIRC characterization will define intrinsic scatter and dependencies on mass, SFR, and metallicity. Compact-source luminosity functions will establish the statistical foundation for future SKA surveys.

\section{Multi-Wavelength Synergies}

Maximizing the scientific return of the SKA continuum observations requires full integration within the multi-wavelength framework of galaxy evolution studies. ALMA molecular-line data (CO, HCN) constrain dense-gas fractions and star-formation efficiency, linking radio emission to gas-regulated SFRs. \emph{JWST} mid-infrared imaging and spectroscopy provide dust-corrected SFR maps and AGN diagnostics at comparable resolution, enabling pixel-scale calibration of the radio--infrared relation. Near-infrared interferometry from VLTI/GRAVITY provides complementary sub-parsec resolution of nuclear stellar and accretion structures, offering independent constraints on black-hole masses and helping to isolate compact AGN components that are unresolved at SKA-Mid scales. Together, these and similar data anchor the radio--IR correlation, tying thermal dust emission and synchrotron radiation to molecular gas content and star-formation efficiency, and enabling decomposition of deviations driven by AGN activity, jets, or cosmic-ray losses. Wide-field optical integral-field spectroscopy from facilities such as the Wide-field Spectroscopic Telescope (WST) will deliver spatially resolved line diagnostics to identify excitation mechanisms, shocks, and outflows, and to separate AGN from stellar photoionization. X-ray data from \textit{Chandra} and \textit{XMM-Newton} trace hot gas in supernova-driven outflows and test radio-based AGN identifications via the radio/X-ray fundamental plane. SKA H\,{\sc i} imaging at matched resolution provides neutral-gas densities for assessing ionization losses and cosmic-ray coupling.

Large-area surveys (EMU, MIGHTEE, LoTSS) extend these analyses to population scales, establishing statistical context and identifying rare objects. Forthcoming facilities such as the Rubin Observatory LSST and the Roman Space Telescope, together with ongoing surveys from \textit{Euclid}, will supply complementary optical--infrared imaging, redshifts, and stellar-mass estimates, enabling unified radio--infrared--optical studies of star formation, dust, and magnetic feedback across environments.

\section{Conclusions}

The SKA-AA4 continuum program combines high sensitivity, sub-arcsecond resolution, and broad frequency coverage to achieve complete separation of thermal and non-thermal radio components from individual star-forming regions to entire galaxies. Its tiered design balances deep physical calibration, population-level validation, and statistical context, linking detailed local studies to large-scale trends. Time-domain observations provide an independent diagnostic of ongoing star formation through the direct detection of supernovae and enable the identification of variable accretion-powered sources.

The resulting data will deliver frequency-dependent, environment-aware radio--SFR calibrations, quantitative constraints on cosmic-ray transport and magnetic-field regulation, and dust-unbiased inventories of compact sources associated with both star formation and accretion. These measurements establish the local-Universe reference needed to interpret the unresolved radio continuum emission of galaxies across cosmic time.

\section*{Acknowledgements}
 JM, AA and MPT acknowledge financial support from the Spanish grant PID2023-147883NB-C21, funded by MCIU/AEI/ 10.13039/501100011033, as well as support through ERDF/EU. JM, AA, MPT, and GL acknowledge financial support from the Severo Ochoa grant CEX2021-001131-S funded by MCIN/AEI/ 10.13039/501100011033. JM  acknowledges the Spanish Prototype of an SRC (SPSRC) service and support funded by the Ministerio de Ciencia, Innovación y Universidades (MICIU), by the Junta de Andalucía, by the European Regional Development Fund (ERDF) and by the European Union NextGenerationEU/PRTR. The SPSRC acknowledges financial support from the Agencia Estatal de Investigación (AEI) through the "Center of Excellence Severo Ochoa" award to the Instituto de Astrofísica de Andalucía (IAA-CSIC) (SEV-2017-0709) and from the grant CEX2021-001131-S funded by MICIU/AEI/ 10.13039/501100011033.

\bibliographystyle{abbrvnat-maxbibnames4}
\bibliography{chapter}

\end{document}